\documentclass[english,aip,amsmath,amssymb,reprint,apl]{revtex4-1}  
\usepackage[T1]{fontenc}
\usepackage[latin9]{inputenc}
\usepackage{textcomp}
\usepackage{amstext}
\usepackage{graphicx}
\usepackage[thickqspace, squaren]{SIunits}
\usepackage{color}

\newcommand{\sub}[1]{_{\mathrm{#1}}}

\begin{document}
\title{A tip-based source of femtosecond electron pulses at 30keV}
\author{Johannes Hoffrogge}
\author{Jan-Paul Stein}
\author{Michael Kr\"uger}
\author{Michael Förster}
\author{Jakob Hammer}
\author{Dominik Ehberger}
\affiliation{Max Planck Institute of Quantum Optics, Hans-Kopfermann-Str.~1~\mbox{, 85748~Garching,  Germany}}
\author{Peter Baum}
\affiliation{Max Planck Institute of Quantum Optics, Hans-Kopfermann-Str.~1~\mbox{, 85748~Garching,  Germany}}
\affiliation{Ludwig-Maximilians-Universit\"at M\"unchen, Am Coulombwall~1~\mbox{, 85748~Garching,  Germany}}
\author{Peter Hommelhoff}
\email{peter.hommelhoff@fau.de}
\affiliation{Max Planck Institute of Quantum Optics, Hans-Kopfermann-Str.~1~\mbox{, 85748~Garching,  Germany}}
\affiliation{Friedrich-Alexander-Universit\"at Erlangen-N\"urnberg, E.-Rommel-Str. 1~\mbox{, 91058~Erlangen,  Germany}}
\date{\today}

\begin{abstract}
We present a nano-scale photoelectron source, optimized towards ultrashort pulse durations and well-suited for time-resolved diffraction experiments. A tungsten tip, mounted in a suppressor-extractor electrode configuration, allows the generation of 30\,keV electron pulses with an estimated pulse duration of 37\,fs at the gun exit. We infer the pulse duration from particle tracking simulations, which are in excellent agreement with experimental measurements of the electron-optical properties of the source. We furthermore demonstrate femtosecond laser-triggered operation. Besides the short electron pulse duration, a tip-based source is expected to feature a large transverse coherence as well as a nanometric emittance.
\end{abstract}

\maketitle

Time-resolved electron diffraction is a powerful tool to follow structural dynamics in space and time, with ``molecular movies'' as the grand goal~\cite{Zewail2006,Sciaini2011,King2005}. Melting processes in metals~\cite{Sciaini2009} 
and structural phase transitions in crystals~\cite{Baum2007b} have been studied with femtosecond electron diffraction. In these experiments, structural dynamics is initiated by laser pulses and electron pulses serve as the probe. An electron kinetic energy of tens of kilo-electron-volts provides short enough de-Broglie wavelengths to resolve atomic distances and changes thereof on the scale of milli-Angstrom. To obtain an instructive diffraction pattern, more than a full unit cell (typically several nanometers for complex materials) needs to be illuminated coherently.
As pulsed electron generation is generally a rather incoherent process, this is a grand challenge in femtosecond diffraction and microscopy. For example, the transverse coherence of flat photocathodes is limited by their source size and electron energy spread \cite{Sciaini2011}, which restricts imaging to full unit cell dimensions below around 1\,nm.

Of further utmost interest is the electron pulse duration delivered by the source, which is the limiting factor for the temporal resolution in a pump-probe experiment.
Experiments that operate with many electrons per pulse are typically limited by space charge broadening of the electron bunch. Coulomb repulsion is most efficient in lengthening the electron pulses within the acceleration region, where the electron cloud is still rather slow. After acceleration to higher velocities, the effects of space charge are much reduced.
Rather complex experimental means such as rf-cavities~\cite{vanOudheusden2010, Gliserin2012},  
reflectrons and ponderomotive light forces~\cite{Baum2007} allow further post-compression of the pulses to smaller durations at the target. Experimentally, 80\,fs pulses have been achieved \cite{vanOudheusden2010}.
Space charge broadening is avoided altogether when operating the electron source in the multi-shot regime, where each electron pulse contains only a few electrons and the final image is constructed from a large number of experimental cycles \cite{Zewail2010, Aidelsburger2010}.  
Typically $10^6$ to $10^8$ electrons are needed for a single image~\cite{King2005,vanOudheusden2010,Sciaini2011}. The effective pulse duration is then given by the arrival time difference of individual electrons at the sample \cite{Baum2009,Zewail2010,Aidelsburger2010}. Its main cause are path length and kinetic energy differences of the individual electrons contributing to the final diffraction image.
Flat single-electron photocathodes, at best technically possible conditions, cannot provide pulses shorter than about 80\,fs \cite{Aidelsburger2010}. Shortening of the pulses and increase of coherence thus pose two of the most urgent challenges towards atomic-scale recording of dynamics with electron diffraction.

Although very different in scope, yet similar in source requirements, are novel electron accelerators such as direct laser accelerators~\cite{Plettner2006} 
and free electron lasers \cite{McNeil2010}. In order to generate the required low emittance beams,
a tip-based setup appears to be the only source to work directly for these applications \cite{Ganter2008}. The device presented in the following thus may represent the low-energy part of an electron gun for these machines.



The central idea of this work is to replace the usually flat cathode with a sharp metal tip with radius of curvature in the range of several tens of nanometers.
A laser pulse triggers photoemission from the localized area of the tip apex while it is biased with a static voltage that lies just below the threshold for field emission.
The first advantage of this geometry is the large acceleration electric field 
that can be generated at the apex of the tip until field emission from its surface sets in at about 2\,GV$/$m \cite{Yanagisawa2010}. 
A high initial field, orders of magnitude larger than in flat emitters (up to 10\,MV$/$m), helps to minimize the acceleration region and therefore the pulse duration.


The second advantage of a tip-based setup is its small (virtual) source size, which is not given by the laser spot size but by the tip size itself. Typical virtual source sizes for field emitter tips lie in the range of one to a few nanometers~\cite{Cho2004}. 
As the transverse coherence length of the electron beam scales inversely proportional to the virtual source size~\cite{Cho2004}, a significant increase in coherence length compared to flat cathode setups is expected. In the ultimate limit of a single emitting atom, even fully coherent electron sources have been demonstrated \cite{Chang2009}.
The nanometric source size furthermore results in extremely low emittance electron beams. For a (comparably large) virtual source size of $5\,$nm and an emission angle of 5°, the emittance of the source presented in the following is as small as $\sim 0.5\,$nm (or 0.0005\,mm\,mrad).
In this work we focus on the temporal aspects of the tip-based electron source, but do expect to observe significant improvements in spatial resolution and emittance in the future as well.

\begin{figure}
  \centerline{\includegraphics[clip,width=85mm]{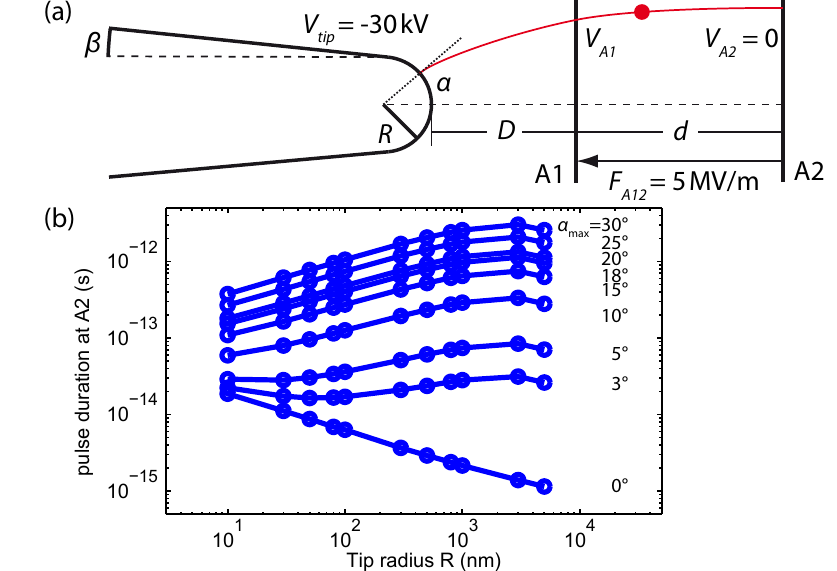}}
  \caption[SetUpTimingJitter]{(Color online) Concept of a tip-based femtosecond electron source.
  a) Sketch of the electrode geometry consisting of a field emission tip placed in front of two anodes.
  The voltage $V_{A1}$ on the first anode and the distances $D$ and $d$ are chosen to satisfy two constraints: an electric field of $F_{tip}=2\,$GV/m at the tip and $F\sub{A12}=5\,$MV/m at the surfaces of the anodes. The trajectory of an electron, emitted under an angle $\alpha$ with respect to the optical axis, is shown in red.
  b) Dependence of the differences in arrival time of emitted electrons on the tip radius $R$ for varying maximum emission angles $\alpha_{max}$. The electrons have an initial energy spread of $\Delta E=0.3\,$eV and their final kinetic energy is 30\,keV.
  }
  \label{fig:SetUpTJ}
\end{figure}

We start with theoretical considerations and present a numerical model of an idealized tip-to-anode geometry. Experimental results from the implemented, slightly modified geometry follow below.
To study the effect of a tip-based setup on the electron pulse duration, we modeled the propagation of single electrons in the geometry depicted in Fig.~\ref{fig:SetUpTJ}\,(a). The final kinetic energy of the electrons is set to $E_{kin} = 30\,$keV, suitable for diffraction from complex materials and molecular systems.
The central concept of our work is to maintain the highest possible electric field along the entire acceleration distance. At the tip, the electric field strength is limited by the onset of field emission to $F\sub{tip} = 2\,$GV/m. At the remaining, macroscopic electrodes, we constrain ourselves to 5\,MV/m, which assures a safety margin with respect to the values commonly reported for polished steel parts~\cite{King2005}.
The requirement to satisfy these boundary conditions for fixed final kinetic energy $E_{kin}$ and arbitrary tip radius $R$ necessitates a setup that incorporates the tip and at least two electrodes, first considered in Ref. \bibpunct{}{}{,}{n}{}{}\cite{Hommelhoff2009}\bibpunct{}{}{,}{s}{}{}.
In the first part of the setup, the distance $D$ between tip and first anode, as well as the voltage difference between those elements, are chosen to fulfill the boundary conditions on both the tip and anode field. The space between the first and second anode then serves as a parallel-plate capacitor with a constant field of $F\sub{A12}=5$\,MV$/$m to accelerate the electrons to the desired final kinetic energy.
All anodes are modeled as infinitely thin sheets that are permeable to electrons. This neglects timing variations arising from the lens action of aperture holes, which in the experiment can be compensated for by additional lens elements forming an overall 'isochronic' lens system \cite{Weninger2012}.

Due to geometric field enhancement, the local electric field at the tip depends on its geometry, which is characterized by the tip radius $R$ and the half opening angle $\beta$ of the tip shaft.
We set $\beta = 3$°, a value that is consistent with the tungsten tips commonly used in our laboratory~\cite{Schenk2010}.
We then performed simulations for various $R$ and scaled both $D$ and $V\sub{A1}$ to yield the above mentioned maximum surface fields.

The timing in a tip-based setup depends on two mechanisms. First, each position on the tip surface can emit electrons with different initial velocities, given by the initial energy distribution of the photoemission process. Second, the inhomogeneity of the acceleration field around the tip produces path length and timing differences between trajectories of electrons that leave the tip under various emission angles $\alpha$.
To simulate the timing of the source, we assume that the electrons are emitted by one-photon photoemission with the central energy of the laser spectrum lying just above the effective surface barrier height of the tip material~\cite{Aidelsburger2010}. The initial electron energy distribution is then assumed to have a width of $\Delta E = 0.3\,$eV (full width at half maximum) \cite{NoteAPL2013}.
The angular dependence of the time of flight differences between individual electrons is studied by considering varying maximum emission angles $\alpha\sub{max}$, which would correspond to the introduction of a clipping aperture with varying opening diameters in the experiment.
We define the pulse duration after the source as the arrival time difference between two extreme electron trajectories. The first, fastest electron trajectory starts on the tip axis with an initial energy of 0.3\,eV, the second, slowest is emitted under an angle $\alpha\sub{max}$ and starts with zero initial energy.
Very recently, Paarmann \textit{et al.}~\cite{Paarmann2012} performed similar simulations 
for single, low energy electrons at $200\,$eV. Apart from the much lower final kinetic energy, they performed their simulations with a tip held at a fixed voltage and placed in front of a single electrode. Their configuration does not yield the highest possible field strengths and therefore cannot produce the minimum achievable pulse duration.

The variation of the electron pulse duration with $R$ and $\alpha_{max}$ in our geometry is shown in Fig.~\ref{fig:SetUpTJ}\,(b).
If we only consider electrons that are emitted on the optical axis ($\alpha_{max}=0$°), the differences in arrival time are solely given by the initial velocity (corresponding to $\Delta E$) and decrease with increasing tip radius. This is because the distance $D$ between tip and first anode increases with $R$, so that the electrons travel a larger fraction of the complete acceleration distance in a region where the electrical field is higher than that in the parallel-plate stage between A1 and A2.
For finite $\alpha_{max}$, the effective path length differs between electrons emitted under different emission angles $\alpha$ and additionally contributes to the pulse duration. Up to $\alpha_{max}\approx5$°, the duration is dominated by $\Delta E$ for small $R$
, so that it initially decreases with increasing $R$. At larger $R$, the effects of the path length differences dominate and the pulse length starts to increase with $R$.
At $\alpha_{max}\approx5$° and for tip radii below $R=30\,$nm, the pulse duration becomes approximately constant. These parameters thus represent a reasonable compromise between a small pulse duration and a large electron current. 
For these settings, the simulations suggest that an effective electron pulse duration of 30\,fs should be feasible in a tip-based setup.
This is a 5-fold improvement compared to a flat cathode geometry operated at 10\,MV$/$m, where the corresponding pulse duration would amount to about 150\,fs at $\Delta E=0.3\,$eV~\cite{Aidelsburger2010}.

\begin{figure}
  \centerline{\includegraphics[clip,width=70mm]{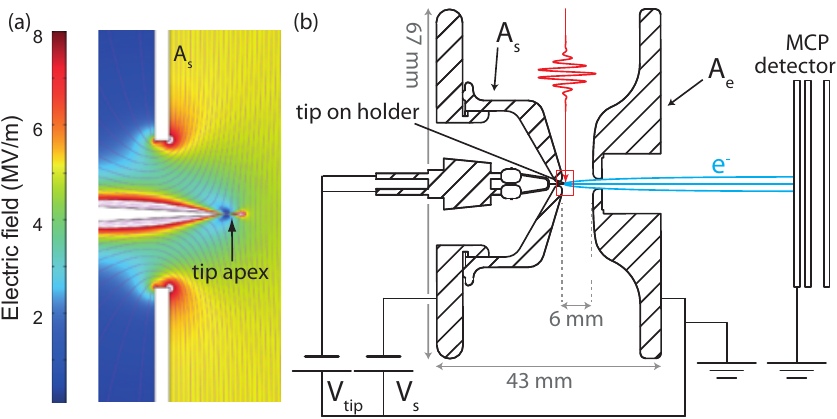}}
  \caption[tipfed_setup]{(Color online) Setup for laser-triggered electron emission.
  a) Simulated electric field magnitude around the tip, which is inserted through a hole in the suppressor $A_s$ with the tip apex placed at 0.5\,mm in front of the electrode surface. The suppressor at $V_s$ is biased with respect to the tip to yield a tip surface field of $F\sub{tip}=2$\,GV/m.
  b) Complete gun setup and experimental arrangement. The geometry provides enough space to focus a triggering laser pulse, shown in red, onto the tip. The electron beam is depicted in blue.
  The grounded extractor $A_e$ provides space for an aperture with a diameter of $94\,\micro\meter$ to limit the maximum emission angle to $\alpha_{max}=5$°.
  For an electron energy of 30\,keV and a tip radius of $R=100\,$nm, we operate the setup at $V_{tip}=-30\,$kV and $V_{s}=-33\,$kV.
}
  \label{fig:tipfed_setup}
\end{figure}

In the experiment, we realized a slightly different electrode geometry that also satisfies the above mentioned boundary conditions and generates high initial acceleration.
The tip apex is sandwiched between a suppressor anode placed about $D=500\micro\meter$ behind the tip apex and an extractor electrode in front of the tip, see Fig.~\ref{fig:tipfed_setup}. The purpose of the suppressor anode is to be experimentally able to reduce $F\sub{tip}$ to the desired value just below the onset of field emission by simply varying the applied voltage $V\sub{sub}$.
The surface fields at the suppressor and extractor are then given by the distance $d$ between suppressor and extractor as well as by the voltage bias of the tip-suppressor unit with respect to the (grounded) extractor.
Compared to the geometry of Fig.~\ref{fig:SetUpTJ}\,(a), this configuration allows the generation of the desired $F\sub{tip}$ for larger distances between tip and suppressor and thus provides better optical access for the laser beam used for photoemission. 
The configuration of Fig.~\ref{fig:tipfed_setup} yields even smaller pulse durations than the geometry of Fig.~\ref{fig:SetUpTJ}\,(a).
This is due to a net focusing of off-axis rays in the field of the plate capacitor formed by suppressor and extractor \cite{Paarmann2012}.
For tip radii of about 100\,nm, the pulse duration at the exit of the extractor electrode amounts to 37\,fs for $\Delta E=0.3$\,eV and $\alpha\sub{max}=5$°.
Considering subsequent field-free propagation, the pulse duration at a distance of 50\,mm from the extractor will be~70\,fs.

\begin{figure}
  \centerline{\includegraphics[clip,width=85mm]{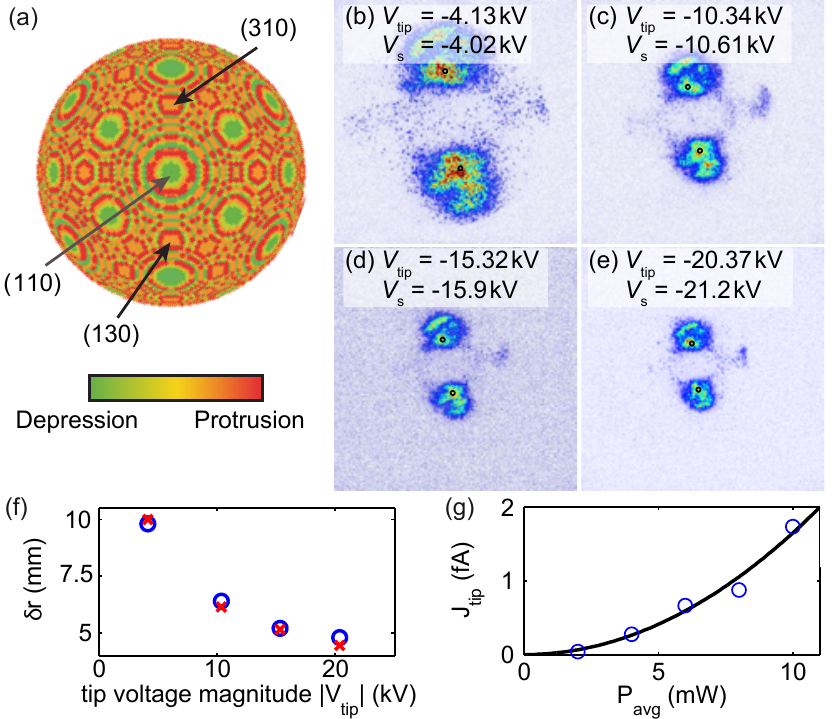}}
  \caption{(Color online) Experimental performance of the tip-based electron gun. (a) Top view of a ball model of the crystal structure at the tip apex. The color encodes the relative distance of the tip surface from the surface of an idealized, hemispherical tip apex.
   The position of the two low work function planes $(310)$ and $(130)$, which predominantly emit electrons in field emission, are highlighted.
  (b)-(e) Field emission patterns for various voltage combinations $V_{tip}$ and $V_s$, measured 72\,mm behind the tip.
  The position of the beamlets that originate from the $\{310\}$ planes are indicated by black circles.
  Here, the kinetic energy is not fixed at 30\,keV but given by $e V_{\mathrm{tip}}$.
  (f) Comparison of the distance between the $\{310\}$ emission spots on the screen (crosses) to that obtained from particle tracking simulation (circles), for a tip with $R = 50$\,nm and $D = 0.75$\,mm.
  (g) Laser-triggered electron emission: Dependence of electron current on the incident laser power together with a quadratic fit.
  }
  \label{fig:focusing_effect}
\end{figure}

We now compare experimental DC field emission patterns of the source for various voltage settings with particle tracking simulations. 
To be able to study the influence of the suppressor electrode on the electron emission from the tip, we employed a tungsten tip with the $[110]$-direction pointing forward, see Fig.~\ref{fig:focusing_effect}\,(a). The whole setup is mounted in a vacuum chamber with a base pressure of $2\times10^{-10}$\,mbar. To avoid electrical breakdown, the suppressor anode is fabricated from stainless steel and polished to optical quality.
We employ an atomically clean tip with a radius of $R\approx 50\,$nm, where residual atoms have been removed by field evaporation during  positively biased operation.
Electron emission from the tip then happens preferentially from the two low-work function planes $\{310\}$, which span an angle of 26.6° with the tip axis, see Fig.~\ref{fig:focusing_effect}\,(a). This allows measuring the magnification of the tip surface on the detector by determining the distance between the two  electron beamlets.
Figs.~\ref{fig:focusing_effect}\,(b)-(e) depict field emission patterns for various tip voltages $V\sub{tip}$. The extractor anode, which would otherwise clip the electron beam, has been removed for these measurements so that the whole emission pattern can be observed at a grounded microchannel plate detector placed 72\,mm away from the tip.
For each of the four tip voltages $V\sub{tip}$, the suppressor voltage $V\sub{s}$ has been adjusted to yield a tip field slightly above field emission threshold. Increasing the magnitude of $V\sub{tip}$ and $V\sub{s}$ then increases the approximately homogeneous field between suppressor and detector. This leads to the observed focusing of the emitted electron beam.
Fig.~\ref{fig:focusing_effect}\,(f) compares the measured distances with results from particle tracking simulations. The excellent agreement indicates that our simulation results, including the values derived for the pulse duration in the full setup, are trustworthy.
The observed effective narrowing of the electron beam at higher $V_{tip}$ also confirms that a reduction of path length differences for off-axis rays is possible by placing the tip in a plate-capacitor setup.
We have successfully generated electron beams with energies up to 30\,keV in field emission while stably operating the critical part of the setup without any surface breakdown.

Femtosecond photoemission has been achieved in the setup of Fig.~\ref{fig:tipfed_setup} with laser pulses from a titanium:sapphire oscillator ($\sim$800\,nm central wavelength,  $\sim$10\,fs pulse duration, 80\,MHz repetition rate). These pulses were focused to a spot radius of $8.5\,\micro\meter$ (1/e$^2$ intensity) at the tip, yielding a peak intensity of $I_{{\rm peak}}\approx 1\times10^{11}$\,W/cm$^{2}$.

%
%
%
The setup was operated at $V_{tip}=-8.00\,$kV and $V_s = -8.25\,$kV, corresponding to $F_{tip}=1.6\,$GV$/$m.
The quadratic dependence of the photocurrent on the laser intensity, shown in Fig.~\ref{fig:focusing_effect}\,(g), indicates a two-photon emission process, as expected for a mean photon energy of about 1.55\,eV and an effective reduction of the work function of tungsten (4.35\,eV) by 1.5\,eV due to the Schottky effect.
The average number of electrons per pulse at the detector was $1.4\times10^{-4}$. The main reason for the low electron number in these measurements was the use of a tip in [110]-orientation. Therefore, the extractor anode clipped most of the emitted electrons that  originate predominantly from off-axis emission sites (compare Figs.~\ref{fig:focusing_effect}\,(b)-(e)).
This will be avoided by using a tip in [310]-orientation in future experiments, where emission is centered around the tip axis.
In a different experiment but using the same laser system, we have observed a maximum electron current of 2000 electrons per pulse in stable emission from a [310]-oriented tip.
We conclude from the emission patterns that about $2\%$ of the emitted electrons will be transmitted when limiting the beam to a maximum emission angle of $\alpha_{max}=5$° to ensure a pulse duration of 37\,fs at the gun exit.
A current of one electron per pulse at the source would then allow the acquisition of a time-resolved diffraction snapshot within minutes using a MHz laser system.

In summary, we have experimentally demonstrated a compact tip-based electron source optimized towards small pulse duration. A tip pointing through one anode and facing another is ideal in terms of pulse length and optical access. We have successfully operated the gun in DC-field emission mode generating 30\,keV electrons. Experimental and simulated emission patterns agree well and strongly support the results of numerical particle tracking simulations. For a 5$^{\circ}$ electron beam acceptance and an initial electron energy width of $\Delta E=0.3$\,eV, a pulse duration of 37\,fs directly at the gun exit and 70\,fs at 5\,cm behind the gun is expected.
First laser-triggered operation of the gun has been demonstrated.

{\bf Acknowledgments}

This work was supported in part by the DFG Cluster of Excellence Munich Center for Advanced Photonics and the DARPA AXiS program. JH and MF acknowledge support from IMPRS-APS. 
PB acknowledges support from ERC and Rudolf-Kaiser-Foundation.
%


\begin{thebibliography}{22}%
\makeatletter
\providecommand \@ifxundefined [1]{%
 \@ifx{#1\undefined}
}%
\providecommand \@ifnum [1]{%
 \ifnum #1\expandafter \@firstoftwo
 \else \expandafter \@secondoftwo
 \fi
}%
\providecommand \@ifx [1]{%
 \ifx #1\expandafter \@firstoftwo
 \else \expandafter \@secondoftwo
 \fi
}%
\providecommand \natexlab [1]{#1}%
\providecommand \enquote  [1]{``#1''}%
\providecommand \bibnamefont  [1]{#1}%
\providecommand \bibfnamefont [1]{#1}%
\providecommand \citenamefont [1]{#1}%
\providecommand \href@noop [0]{\@secondoftwo}%
\providecommand \href [0]{\begingroup \@sanitize@url \@href}%
\providecommand \@href[1]{\@@startlink{#1}\@@href}%
\providecommand \@@href[1]{\endgroup#1\@@endlink}%
\providecommand \@sanitize@url [0]{\catcode `\\12\catcode `\$12\catcode
  `\&12\catcode `\#12\catcode `\^12\catcode `\_12\catcode `\%12\relax}%
\providecommand \@@startlink[1]{}%
\providecommand \@@endlink[0]{}%
\providecommand \url  [0]{\begingroup\@sanitize@url \@url }%
\providecommand \@url [1]{\endgroup\@href {#1}{\urlprefix }}%
\providecommand \urlprefix  [0]{URL }%
\providecommand \Eprint [0]{\href }%
\providecommand \doibase [0]{http://dx.doi.org/}%
\providecommand \selectlanguage [0]{\@gobble}%
\providecommand \bibinfo  [0]{\@secondoftwo}%
\providecommand \bibfield  [0]{\@secondoftwo}%
\providecommand \translation [1]{[#1]}%
\providecommand \BibitemOpen [0]{}%
\providecommand \bibitemStop [0]{}%
\providecommand \bibitemNoStop [0]{.\EOS\space}%
\providecommand \EOS [0]{\spacefactor3000\relax}%
\providecommand \BibitemShut  [1]{\csname bibitem#1\endcsname}%
\let\auto@bib@innerbib\@empty
\bibitem [{\citenamefont {Zewail}(2006)}]{Zewail2006}%
  \BibitemOpen
  \bibfield  {author} {\bibinfo {author} {\bibfnamefont {A.~H.}\ \bibnamefont
  {Zewail}},\ }\href@noop {} {\bibfield  {journal} {\bibinfo  {journal} {Annu.
  Rev. Phys. Chem.}\ }\textbf {\bibinfo {volume} {57}},\ \bibinfo {pages} {65}
  (\bibinfo {year} {2006})}\BibitemShut {NoStop}%
\bibitem [{\citenamefont {Sciaini}\ and\ \citenamefont
  {Miller}(2011)}]{Sciaini2011}%
  \BibitemOpen
  \bibfield  {author} {\bibinfo {author} {\bibfnamefont {G.}~\bibnamefont
  {Sciaini}}\ and\ \bibinfo {author} {\bibfnamefont {R.~J.~D.}\ \bibnamefont
  {Miller}},\ }\href@noop {} {\bibfield  {journal} {\bibinfo  {journal} {Rep.
  Prog. Phys.}\ }\textbf {\bibinfo {volume} {74}},\ \bibinfo {pages} {096101}
  (\bibinfo {year} {2011})}\BibitemShut {NoStop}%
\bibitem [{\citenamefont {King}\ \emph {et~al.}(2005)\citenamefont {King},
  \citenamefont {Campbell}, \citenamefont {Frank}, \citenamefont {Reed},
  \citenamefont {Schmerge}, \citenamefont {Siwick}, \citenamefont {Stuart},\
  and\ \citenamefont {Weber}}]{King2005}%
  \BibitemOpen
  \bibfield  {author} {\bibinfo {author} {\bibfnamefont {W.~E.}\ \bibnamefont
  {King}}, \bibinfo {author} {\bibfnamefont {G.~H.}\ \bibnamefont {Campbell}},
  \bibinfo {author} {\bibfnamefont {A.}~\bibnamefont {Frank}}, \bibinfo
  {author} {\bibfnamefont {B.}~\bibnamefont {Reed}}, \bibinfo {author}
  {\bibfnamefont {J.~F.}\ \bibnamefont {Schmerge}}, \bibinfo {author}
  {\bibfnamefont {B.~J.}\ \bibnamefont {Siwick}}, \bibinfo {author}
  {\bibfnamefont {B.~C.}\ \bibnamefont {Stuart}}, \ and\ \bibinfo {author}
  {\bibfnamefont {P.~M.}\ \bibnamefont {Weber}},\ }\href@noop {} {\bibfield
  {journal} {\bibinfo  {journal} {J.~Appl.~Phys.}\ }\textbf {\bibinfo {volume}
  {97}},\ \bibinfo {eid} {111101} (\bibinfo {year} {2005})}\BibitemShut
  {NoStop}%
\bibitem [{\citenamefont {Sciaini}\ \emph {et~al.}(2009)\citenamefont
  {Sciaini}, \citenamefont {Harb}, \citenamefont {Kruglik}, \citenamefont
  {Payer}, \citenamefont {Hebeisen}, \citenamefont {Heringdorf}, \citenamefont
  {Yamaguchi}, \citenamefont {Hoegen}, \citenamefont {Ernstorfer},\ and\
  \citenamefont {Miller}}]{Sciaini2009}%
  \BibitemOpen
  \bibfield  {author} {\bibinfo {author} {\bibfnamefont {G.}~\bibnamefont
  {Sciaini}}, \bibinfo {author} {\bibfnamefont {M.}~\bibnamefont {Harb}},
  \bibinfo {author} {\bibfnamefont {S.~G.}\ \bibnamefont {Kruglik}}, \bibinfo
  {author} {\bibfnamefont {T.}~\bibnamefont {Payer}}, \bibinfo {author}
  {\bibfnamefont {C.~T.}\ \bibnamefont {Hebeisen}}, \bibinfo {author}
  {\bibfnamefont {F.-J. M.~z.}\ \bibnamefont {Heringdorf}}, \bibinfo {author}
  {\bibfnamefont {M.}~\bibnamefont {Yamaguchi}}, \bibinfo {author}
  {\bibfnamefont {M.~H.-v.}\ \bibnamefont {Hoegen}}, \bibinfo {author}
  {\bibfnamefont {R.}~\bibnamefont {Ernstorfer}}, \ and\ \bibinfo {author}
  {\bibfnamefont {R.~J.~D.}\ \bibnamefont {Miller}},\ }\href@noop {} {\bibfield
   {journal} {\bibinfo  {journal} {Nature}\ }\textbf {\bibinfo {volume}
  {458}},\ \bibinfo {pages} {56} (\bibinfo {year} {2009})}\BibitemShut
  {NoStop}%
\bibitem [{\citenamefont {Baum}\ \emph {et~al.}(2007)\citenamefont {Baum},
  \citenamefont {Yang},\ and\ \citenamefont {Zewail}}]{Baum2007b}%
  \BibitemOpen
  \bibfield  {author} {\bibinfo {author} {\bibfnamefont {P.}~\bibnamefont
  {Baum}}, \bibinfo {author} {\bibfnamefont {D.-S.}\ \bibnamefont {Yang}}, \
  and\ \bibinfo {author} {\bibfnamefont {A.~H.}\ \bibnamefont {Zewail}},\
  }\href@noop {} {\bibfield  {journal} {\bibinfo  {journal} {Science}\ }\textbf
  {\bibinfo {volume} {318}},\ \bibinfo {pages} {788} (\bibinfo {year}
  {2007})}\BibitemShut {NoStop}%
\bibitem [{\citenamefont {van Oudheusden}\ \emph {et~al.}(2010)\citenamefont
  {van Oudheusden}, \citenamefont {Pasmans}, \citenamefont {van~der Geer},
  \citenamefont {de~Loos}, \citenamefont {van~der Wiel},\ and\ \citenamefont
  {Luiten}}]{vanOudheusden2010}%
  \BibitemOpen
  \bibfield  {author} {\bibinfo {author} {\bibfnamefont {T.}~\bibnamefont {van
  Oudheusden}}, \bibinfo {author} {\bibfnamefont {P.~L. E.~M.}\ \bibnamefont
  {Pasmans}}, \bibinfo {author} {\bibfnamefont {S.~B.}\ \bibnamefont {van~der
  Geer}}, \bibinfo {author} {\bibfnamefont {M.~J.}\ \bibnamefont {de~Loos}},
  \bibinfo {author} {\bibfnamefont {M.~J.}\ \bibnamefont {van~der Wiel}}, \
  and\ \bibinfo {author} {\bibfnamefont {O.~J.}\ \bibnamefont {Luiten}},\
  }\href {\doibase 10.1103/PhysRevLett.105.264801} {\bibfield  {journal}
  {\bibinfo  {journal} {Phys.~Rev.~Lett.}\ }\textbf {\bibinfo {volume} {105}},\
  \bibinfo {pages} {264801} (\bibinfo {year} {2010})}\BibitemShut {NoStop}%
\bibitem [{\citenamefont {{Gliserin}}\ \emph {et~al.}(2012)\citenamefont
  {{Gliserin}}, \citenamefont {{Apolonski}}, \citenamefont {{Krausz}},\ and\
  \citenamefont {{Baum}}}]{Gliserin2012}%
  \BibitemOpen
  \bibfield  {author} {\bibinfo {author} {\bibfnamefont {A.}~\bibnamefont
  {{Gliserin}}}, \bibinfo {author} {\bibfnamefont {A.}~\bibnamefont
  {{Apolonski}}}, \bibinfo {author} {\bibfnamefont {F.}~\bibnamefont
  {{Krausz}}}, \ and\ \bibinfo {author} {\bibfnamefont {P.}~\bibnamefont
  {{Baum}}},\ }\href {\doibase 10.1088/1367-2630/14/7/073055} {\bibfield
  {journal} {\bibinfo  {journal} {New Journal of Physics}\ }\textbf {\bibinfo
  {volume} {14}},\ \bibinfo {pages} {073055} (\bibinfo {year}
  {2012})}\BibitemShut {NoStop}%
\bibitem [{\citenamefont {Baum}\ and\ \citenamefont {Zewail}(2007)}]{Baum2007}%
  \BibitemOpen
  \bibfield  {author} {\bibinfo {author} {\bibfnamefont {P.}~\bibnamefont
  {Baum}}\ and\ \bibinfo {author} {\bibfnamefont {A.~H.}\ \bibnamefont
  {Zewail}},\ }\href@noop {} {\bibfield  {journal} {\bibinfo  {journal} {PNAS}\
  }\textbf {\bibinfo {volume} {104}},\ \bibinfo {pages} {18409} (\bibinfo
  {year} {2007})}\BibitemShut {NoStop}%
\bibitem [{\citenamefont {Zewail}(2010)}]{Zewail2010}%
  \BibitemOpen
  \bibfield  {author} {\bibinfo {author} {\bibfnamefont {A.~H.}\ \bibnamefont
  {Zewail}},\ }\href@noop {} {\bibfield  {journal} {\bibinfo  {journal}
  {Science}\ }\textbf {\bibinfo {volume} {328}},\ \bibinfo {pages} {187}
  (\bibinfo {year} {2010})}\BibitemShut {NoStop}%
\bibitem [{\citenamefont {Aidelsburger}\ \emph {et~al.}(2010)\citenamefont
  {Aidelsburger}, \citenamefont {Kirchner}, \citenamefont {Krausz},\ and\
  \citenamefont {Baum}}]{Aidelsburger2010}%
  \BibitemOpen
  \bibfield  {author} {\bibinfo {author} {\bibfnamefont {M.}~\bibnamefont
  {Aidelsburger}}, \bibinfo {author} {\bibfnamefont {F.~O.}\ \bibnamefont
  {Kirchner}}, \bibinfo {author} {\bibfnamefont {F.}~\bibnamefont {Krausz}}, \
  and\ \bibinfo {author} {\bibfnamefont {P.}~\bibnamefont {Baum}},\ }\href@noop
  {} {\bibfield  {journal} {\bibinfo  {journal} {PNAS}\ }\textbf {\bibinfo
  {volume} {107}},\ \bibinfo {pages} {19714} (\bibinfo {year}
  {2010})}\BibitemShut {NoStop}%
\bibitem [{\citenamefont {{Baum}}\ and\ \citenamefont
  {{Zewail}}(2009)}]{Baum2009}%
  \BibitemOpen
  \bibfield  {author} {\bibinfo {author} {\bibfnamefont {P.}~\bibnamefont
  {{Baum}}}\ and\ \bibinfo {author} {\bibfnamefont {A.~H.}\ \bibnamefont
  {{Zewail}}},\ }\href {\doibase 10.1016/j.chemphys.2009.07.013} {\bibfield
  {journal} {\bibinfo  {journal} {Chemical Physics}\ }\textbf {\bibinfo
  {volume} {366}},\ \bibinfo {pages} {2} (\bibinfo {year} {2009})}\BibitemShut
  {NoStop}%
\bibitem [{\citenamefont {Plettner}\ \emph {et~al.}(2006)\citenamefont
  {Plettner}, \citenamefont {Lu},\ and\ \citenamefont {Byer}}]{Plettner2006}%
  \BibitemOpen
  \bibfield  {author} {\bibinfo {author} {\bibfnamefont {T.}~\bibnamefont
  {Plettner}}, \bibinfo {author} {\bibfnamefont {P.~P.}\ \bibnamefont {Lu}}, \
  and\ \bibinfo {author} {\bibfnamefont {R.~L.}\ \bibnamefont {Byer}},\
  }\href@noop {} {\bibfield  {journal} {\bibinfo  {journal} {PRSTAB}\ }\textbf
  {\bibinfo {volume} {9}},\ \bibinfo {eid} {111301} (\bibinfo {year}
  {2006})}\BibitemShut {NoStop}%
\bibitem [{\citenamefont {{McNeil}}\ and\ \citenamefont
  {{Thompson}}(2010)}]{McNeil2010}%
  \BibitemOpen
  \bibfield  {author} {\bibinfo {author} {\bibfnamefont {B.~W.~J.}\
  \bibnamefont {{McNeil}}}\ and\ \bibinfo {author} {\bibfnamefont {N.~R.}\
  \bibnamefont {{Thompson}}},\ }\href {\doibase 10.1038/nphoton.2010.239}
  {\bibfield  {journal} {\bibinfo  {journal} {Nat. Phot.}\ }\textbf {\bibinfo
  {volume} {4}},\ \bibinfo {pages} {814} (\bibinfo {year} {2010})}\BibitemShut
  {NoStop}%
\bibitem [{\citenamefont {Ganter}\ \emph {et~al.}(2008)\citenamefont {Ganter},
  \citenamefont {Bakker}, \citenamefont {Gough}, \citenamefont {Leemann},
  \citenamefont {Paraliev}, \citenamefont {Pedrozzi}, \citenamefont
  {Le~Pimpec}, \citenamefont {Schlott}, \citenamefont {Rivkin},\ and\
  \citenamefont {A.}}]{Ganter2008}%
  \BibitemOpen
  \bibfield  {author} {\bibinfo {author} {\bibfnamefont {R.}~\bibnamefont
  {Ganter}}, \bibinfo {author} {\bibfnamefont {R.}~\bibnamefont {Bakker}},
  \bibinfo {author} {\bibfnamefont {C.}~\bibnamefont {Gough}}, \bibinfo
  {author} {\bibfnamefont {S.~C.}\ \bibnamefont {Leemann}}, \bibinfo {author}
  {\bibfnamefont {M.}~\bibnamefont {Paraliev}}, \bibinfo {author}
  {\bibfnamefont {M.}~\bibnamefont {Pedrozzi}}, \bibinfo {author}
  {\bibfnamefont {F.}~\bibnamefont {Le~Pimpec}}, \bibinfo {author}
  {\bibfnamefont {V.}~\bibnamefont {Schlott}}, \bibinfo {author} {\bibfnamefont
  {L.}~\bibnamefont {Rivkin}}, \ and\ \bibinfo {author} {\bibfnamefont
  {W.}~\bibnamefont {A.}},\ }\href {\doibase 10.1103/PhysRevLett.100.064801}
  {\bibfield  {journal} {\bibinfo  {journal} {Phys.~Rev.~Lett.}\ }\textbf
  {\bibinfo {volume} {100}},\ \bibinfo {pages} {064801} (\bibinfo {year}
  {2008})}\BibitemShut {NoStop}%
\bibitem [{\citenamefont {Yanagisawa}\ \emph {et~al.}(2010)\citenamefont
  {Yanagisawa}, \citenamefont {Hafner}, \citenamefont {Doná}, \citenamefont
  {Kl{\"o}ckner}, \citenamefont {Leuenberger}, \citenamefont {Greber},
  \citenamefont {Osterwalder},\ and\ \citenamefont
  {Hengsberger}}]{Yanagisawa2010}%
  \BibitemOpen
  \bibfield  {author} {\bibinfo {author} {\bibfnamefont {H.}~\bibnamefont
  {Yanagisawa}}, \bibinfo {author} {\bibfnamefont {C.}~\bibnamefont {Hafner}},
  \bibinfo {author} {\bibfnamefont {P.}~\bibnamefont {Doná}}, \bibinfo {author}
  {\bibfnamefont {M.}~\bibnamefont {Kl{\"o}ckner}}, \bibinfo {author}
  {\bibfnamefont {D.}~\bibnamefont {Leuenberger}}, \bibinfo {author}
  {\bibfnamefont {T.}~\bibnamefont {Greber}}, \bibinfo {author} {\bibfnamefont
  {J.}~\bibnamefont {Osterwalder}}, \ and\ \bibinfo {author} {\bibfnamefont
  {M.}~\bibnamefont {Hengsberger}},\ }\href {\doibase
  10.1103/PhysRevB.81.115429} {\bibfield  {journal} {\bibinfo  {journal}
  {Phys.~Rev.~B}\ }\textbf {\bibinfo {volume} {81}},\ \bibinfo {pages} {115429}
  (\bibinfo {year} {2010})}\BibitemShut {NoStop}%
\bibitem [{\citenamefont {Cho}\ \emph {et~al.}(2004)\citenamefont {Cho},
  \citenamefont {Ichimura}, \citenamefont {Shimizu},\ and\ \citenamefont
  {Oshima}}]{Cho2004}%
  \BibitemOpen
  \bibfield  {author} {\bibinfo {author} {\bibfnamefont {B.}~\bibnamefont
  {Cho}}, \bibinfo {author} {\bibfnamefont {T.}~\bibnamefont {Ichimura}},
  \bibinfo {author} {\bibfnamefont {R.}~\bibnamefont {Shimizu}}, \ and\
  \bibinfo {author} {\bibfnamefont {C.}~\bibnamefont {Oshima}},\ }\href@noop {}
  {\bibfield  {journal} {\bibinfo  {journal} {Phys.~Rev.~Lett.}\ }\textbf
  {\bibinfo {volume} {92}},\ \bibinfo {pages} {246103} (\bibinfo {year}
  {2004})}\BibitemShut {NoStop}%
\bibitem [{\citenamefont {Chang}\ \emph {et~al.}(2009)\citenamefont {Chang},
  \citenamefont {Kuo}, \citenamefont {Hwang},\ and\ \citenamefont
  {Tsong}}]{Chang2009}%
  \BibitemOpen
  \bibfield  {author} {\bibinfo {author} {\bibfnamefont {C.-C.}\ \bibnamefont
  {Chang}}, \bibinfo {author} {\bibfnamefont {H.-S.}\ \bibnamefont {Kuo}},
  \bibinfo {author} {\bibfnamefont {I.-S.}\ \bibnamefont {Hwang}}, \ and\
  \bibinfo {author} {\bibfnamefont {T.~T.}\ \bibnamefont {Tsong}},\ }\href@noop
  {} {\bibfield  {journal} {\bibinfo  {journal} {Nanotech.}\ }\textbf {\bibinfo
  {volume} {20}},\ \bibinfo {pages} {115401} (\bibinfo {year}
  {2009})}\BibitemShut {NoStop}%
\bibitem [{\citenamefont {Hommelhoff}\ \emph {et~al.}(2009)\citenamefont
  {Hommelhoff}, \citenamefont {Kealhofer}, \citenamefont {Aghajani-Talesh},
  \citenamefont {Sortais}, \citenamefont {Foreman},\ and\ \citenamefont
  {Kasevich}}]{Hommelhoff2009}%
  \BibitemOpen
  \bibfield  {author} {\bibinfo {author} {\bibfnamefont {P.}~\bibnamefont
  {Hommelhoff}}, \bibinfo {author} {\bibfnamefont {C.}~\bibnamefont
  {Kealhofer}}, \bibinfo {author} {\bibfnamefont {A.}~\bibnamefont
  {Aghajani-Talesh}}, \bibinfo {author} {\bibfnamefont {Y.~R.}\ \bibnamefont
  {Sortais}}, \bibinfo {author} {\bibfnamefont {S.~M.}\ \bibnamefont
  {Foreman}}, \ and\ \bibinfo {author} {\bibfnamefont {M.~A.}\ \bibnamefont
  {Kasevich}},\ }\href {\doibase 10.1016/j.ultramic.2008.10.021} {\bibfield
  {journal} {\bibinfo  {journal} {Ultramicr.}\ }\textbf {\bibinfo {volume}
  {109}},\ \bibinfo {pages} {423 } (\bibinfo {year} {2009})}\BibitemShut
  {NoStop}%
\bibitem [{\citenamefont {Weninger}\ and\ \citenamefont
  {Baum}(2012)}]{Weninger2012}%
  \BibitemOpen
  \bibfield  {author} {\bibinfo {author} {\bibfnamefont {C.}~\bibnamefont
  {Weninger}}\ and\ \bibinfo {author} {\bibfnamefont {P.}~\bibnamefont
  {Baum}},\ }\href {\doibase 10.1016/j.ultramic.2011.11.018} {\bibfield
  {journal} {\bibinfo  {journal} {Ultramicr.}\ }\textbf {\bibinfo {volume}
  {113}},\ \bibinfo {pages} {145 } (\bibinfo {year} {2012})}\BibitemShut
  {NoStop}%
\bibitem [{\citenamefont {Schenk}\ \emph {et~al.}(2010)\citenamefont {Schenk},
  \citenamefont {Kr\"uger},\ and\ \citenamefont {Hommelhoff}}]{Schenk2010}%
  \BibitemOpen
  \bibfield  {author} {\bibinfo {author} {\bibfnamefont {M.}~\bibnamefont
  {Schenk}}, \bibinfo {author} {\bibfnamefont {M.}~\bibnamefont {Kr\"uger}}, \
  and\ \bibinfo {author} {\bibfnamefont {P.}~\bibnamefont {Hommelhoff}},\
  }\href {\doibase 10.1103/PhysRevLett.105.257601} {\bibfield  {journal}
  {\bibinfo  {journal} {Phys.~Rev.~Lett.}\ }\textbf {\bibinfo {volume} {105}},\
  \bibinfo {pages} {257601} (\bibinfo {year} {2010})}\BibitemShut {NoStop}%
\bibitem [{\citenamefont {{see Suppl. Material for details}}()}]{NoteAPL2013}%
  \BibitemOpen
  \bibfield  {author} {\bibinfo {author} {\bibnamefont {{see Suppl. Material
  for details}}}\ }\href@noop {} {\ }\BibitemShut {NoStop}%
\bibitem [{\citenamefont {{Paarmann}}\ \emph {et~al.}(2012)\citenamefont
  {{Paarmann}}, \citenamefont {{Gulde}}, \citenamefont {{M{\"u}ller}},
  \citenamefont {{Sch{\"a}fer}}, \citenamefont {{Schweda}}, \citenamefont
  {{Maiti}}, \citenamefont {{Xu}}, \citenamefont {{Hohage}}, \citenamefont
  {{Schenk}}, \citenamefont {{Ropers}},\ and\ \citenamefont
  {{Ernstorfer}}}]{Paarmann2012}%
  \BibitemOpen
  \bibfield  {author} {\bibinfo {author} {\bibfnamefont {A.}~\bibnamefont
  {{Paarmann}}}, \bibinfo {author} {\bibfnamefont {M.}~\bibnamefont {{Gulde}}},
  \bibinfo {author} {\bibfnamefont {M.}~\bibnamefont {{M{\"u}ller}}}, \bibinfo
  {author} {\bibfnamefont {S.}~\bibnamefont {{Sch{\"a}fer}}}, \bibinfo {author}
  {\bibfnamefont {S.}~\bibnamefont {{Schweda}}}, \bibinfo {author}
  {\bibfnamefont {M.}~\bibnamefont {{Maiti}}}, \bibinfo {author} {\bibfnamefont
  {C.}~\bibnamefont {{Xu}}}, \bibinfo {author} {\bibfnamefont {T.}~\bibnamefont
  {{Hohage}}}, \bibinfo {author} {\bibfnamefont {F.}~\bibnamefont {{Schenk}}},
  \bibinfo {author} {\bibfnamefont {C.}~\bibnamefont {{Ropers}}}, \ and\
  \bibinfo {author} {\bibfnamefont {R.}~\bibnamefont {{Ernstorfer}}},\ }\href
  {\doibase 10.1063/1.4768204} {\bibfield  {journal} {\bibinfo  {journal}
  {J.~Appl.~Phys.}\ }\textbf {\bibinfo {volume} {112}},\ \bibinfo {pages}
  {113109} (\bibinfo {year} {2012})}\BibitemShut {NoStop}%
\end{thebibliography}

%

\end{document}